\documentclass[11pt]{article}
\usepackage{color,graphicx}
\usepackage[utf8]{inputenc}

\newcommand{\beq}{\begin{equation}}
\newcommand{\eeq}{\end{equation}}
\newcommand{\beqs}{\begin{eqnarray}}
\newcommand{\eeqs}{\end{eqnarray}}

\def \be {\begin{equation}} 
\def \ee {\end{equation}}
\def \bea {\begin{eqnarray}}
\def \eea {\end{eqnarray}}
\def \nn {\nonumber}

\def \lab #1 {\label{#1}}

\def \as {{\alpha_s}}
\def \ra {\rightarrow}

\definecolor{agr}{rgb}{0.0,0.4,0.0}
\definecolor{apu}{rgb}{0.3,0.2,0.6}
\definecolor{dre}{rgb}{0.6,0.0,0.1}

\ifx\pdfoutput\undefined
\input epsf

\def\figscale#1#2{\epsfxsize=#2\epsfbox{#1.eps}}
\else

\def\figscale#1#2{\pdfximage width#2 {#1.pdf}\pdfrefximage\pdflastximage}
\fi

\def \bea{\begin{eqnarray}}
\def \eea{\end{eqnarray}}
\def \ba{\begin{eqnarray}}
\def \ea{\end{eqnarray}}

\def\figscale#1#2{\pdfximage width#2 {#1.pdf}\pdfrefximage\pdflastximage}

\def \nn{\nonumber}

\begin{document}

\begin{center}
\Large
{\bf QCD for electroweak precision measurements: Foundations}\footnote{Submitted for publication by Elsevier Ltd. in a Physics Reports collection of papers on Electroweak Precision Physics, edited by F. Bedeschi, A. Kotwal, M. Ramsey-Musolf, C. Vellidis, and D. Wackeroth.}
\end{center}

\medskip

\begin{center}
{ {\bf George Sterman}\\[5mm]
C.N. Yang Institute for Theoretical Physics and Department of Physics and Astronomy\\
Stony Brook University, Stony Brook NY 11794-3840 USA\\
george.sterman@stonybrook.edu
}
\end{center}

\medskip

\begin{abstract}
The couplings of the strong to electroweak sectors of the Standard Model enable the exploration of each  using our growing knowledge of the other.
In this review, we will follow the sweep of history.  
Starting with QED as a precision theory, 
deep inelastic scattering served as a gateway to the strong interactions,  followed by leptonic annihilation and
quark-antiquark annihilation
in hadron-hadron scattering.  In turn, the resulting understanding
of QCD helped  establish the Standard Model. The same techniques
form the basis for  many precision electroweak measurements at high energy and searches for signs of new physics.
\end{abstract}

\tableofcontents

\section{Introduction}

The era of precision in quantum field theory begins with quantum electrodynamics, the dawn of whose combination
of extensive fields and localized particles  
may be traced back to two discoveries by Heinrich Hertz in the late nineteenth century: electromagnetic waves and the photoelectric effect.
The  need to combine these two concepts was enunciated by Einstein in his identification of electromagnetic
quanta, our photons, as the necessarily {\it local} explanation of photoelectric phenomena.  For more on
these and related historical references below, see Ref.\  \cite{Pais:1986nu}.

With the invention of quantum mechanics came its application to electromagnetism, whose dynamics 
became describable in terms of electromagnetic and Dirac fields, 
filling the universe, yet whose time evolution can be described as a succession of
collections of photons and electrons, states succeeding each other with changing 
particle content.  
At the outset, the theory posed two fundamental problems:  how to deal with the infinite set
of states at very high energy, the ultraviolet problem, and how to treat the influence of
classical radiation on quantum transitions,
 the infrared problem.  

The solution to
the ultraviolet problem, of course is renormalization, in which divergent summations over high energy intermediate
states are absorbed into numbers taken from experiment, in QED the value of the fine-structure
constant $\alpha$ and of the electron mass.  For the QED infrared problem,
it was found adequate to make cross sections or transition probabilities  sufficiently {\it inclusive} in soft radiation \cite{Bloch:1937pw}.   This leads to the naively paradoxical 
result that summing over indefinite numbers of radiated soft photons makes Compton
or Bhabha scattering well-approximated by their lowest-order cross sections at moderate energies .   In a sense,
 final states with soft photons ``disappear" in the inclusive cross section.\footnote{The effects of higher-order QED
 corrections have been reviewed recently in \cite{Afanasev:2023gev}.  An approach combining QED corrections the factorization formalism for QCD described in this document
 has appeared very receently \cite{Cammarota:2025jyr}.}

These powerful ideas were in place for QED by mid twentieth century, but through the 1950s and 1960s it was
not yet known how they could be applied to the strong interactions.   During this period, excited states and what we now term flavor symmetries
were discovered, 
and the concept of quarks was born, but it was far from clear whether
a theory based on quark or hypothetical ``gluon" quantum fields could succeed.

At the same time, however, the hadrons, whatever they were, had both electromagnetic and
weak interactions.   The next route to discovery leveraged electroweak interactions to explore
the structure of hadrons \cite{Panofsky:1968ka} 
and before long  led to quantum chromodynamics, making
the concept of precision applicable to the strong interactions.   Eventually, the
strong interactions repaid the complement, providing new
routes to particle production and precision
measurements in the electroweak sectors of the Standard Model.

In the following, we will review the primary links between strong and electroweak interactions that can be realized
in high energy accelerators, organized around the parton model and the three channels of electron-quark scattering.
The main ideas will be reviewed in the context of deep inelastic scattering (DIS).
These experiments,  their motivation and their catalytic observation of  scaling
\cite{Bloom:1969kc,Breidenbach:1969kd}
 are described in Sec.\ \ref{sec:j-to-partons},
along with some basic features of the parton model \cite{Feynman:1969ej,Bjorken:1969ja}.
Section \ref{sec:partons-to-qcd} recalls how
electromagnetic scattering led to the discovery of quantum chromodynamics (QCD).  In this context,
we develop the concepts of parton distributions, collinear factorization and evolution.   In Sec.\ \ref{sec:fact-to-evol}, we
review the application of perturbation theory in infrared-regulated QCD to physical cross sections,
and the general relationship between factorization and evolution.   In Sec.\ \ref{sec:crossed},
we discuss the crossed channels of lepton-quark scattering, beginning with 
leptonic annihilation to quarks, single-particle inclusive cross sections and the
concept of jet cross sections, and go on to Drell-Yan and allied processes,
describing its collinear factorization properties, and the extension to resummation of
transverse momentum.

The material covered in this article appears for the most part in advanced texts on the
Standard Model in quantum field theory, including \cite{Quigg:2013ufa,Schwartz:2014sze} and in
\cite{CTEQ:1993hwr,Ellis:1996mzs,Collins:2011zzd,Campbell:2017hsr} on perturbative quantum chromodynamics.  The discussion below is
intended to complement these rigorous treatments with a compact presentation
of foundational results supported by heuristic explanations.

\section{From currents to partons}
\label{sec:j-to-partons}

\subsection{Local currents as the bridge}

Even without knowledge of the strong interactions, it is a natural assumption 
that the electroweak window to proton substructure is through
local currents.   By the advent of GeV (then BeV) energy accelerators, this idea was
well-established through QED and Fermi's four-fermion Hamiltonian for the weak interactions.
In QED, photons are emitted and absorbed locally through the term in the Lagrangian that
couples the photon field $A^\mu$ to the electromagnetic current $j_\mu = \bar \psi_e \gamma_\mu \psi_e$.
We then consider the scattering of an electron of momentum $k$ and spin $\lambda$ with a nucleon (proton or neutron)
of momentum $p$ and spin $\sigma$ through the exchange of a single photon, of spacelike momentum $q$,
$q^2=-Q^2<0$, resulting in a single outgoing electron of momentum $k-q$, and some hadronic state $X(p+q)$,
\bea
e(k,\lambda) + N(p,\sigma) \rightarrow e(k-q,\lambda') + X(p+q)\, .
\eea
Such a process is called deep-inelastic if both $Q^2$ and $(p+q)^2$ are
substantially larger than the mass of the proton.   Experiments of this sort became possible 
by the late 1960s, with the development of a multi-GeV linear accelerator for electrons at SLAC.
They continued through experiments at the Tevatron, DESY, CERN and Jefferson Laboratory,
with an extensive future program planned for the Electron-Ion Collider at Brookhaven Laboratory.
They provide the bedrock of our information on the elementary structure of strongly-interacting matter \cite{Blumlein:2023aso}.

At lowest order in QED, the amplitude for such a process can be written as
\bea
A_{eN\rightarrow eX}(\lambda,\lambda',\sigma;q)
&=& \bar{u}(k',\lambda')(-ie\gamma_\mu)u(k,\lambda)
\, \frac{-ig^{\mu\mu'}}{q^2}
\, \langle X|\, -iJ_{\mu'}^{\rm EM}(0)\, |p,\sigma\rangle\, ,
\label{eq:ampl}
\eea
where here $e$ represents the positron charge.
At this stage, and indeed at the time of the original experiments, no assumption is made about what fields might make up
 the hadronic electromagnetic current $J_\mu^{\rm EM}$;  correspondingly, its matrix elements are completely unknown, and in any case
 must depend on the detailed nature of the hadronic final state $X$, which in general consists of many particles. 
 We will see, however, that summing over all possible final states at a given $q$ will simplify the analysis greatly,
 in a manner reminiscent of the treatment of infrared divergences in QED, relating a sum of inelastic processes
 to an elastic process, up to calculable corrections.

Averaging, for simplicity over incoming spins, summing over $\lambda'$ and final states $X(p+q)$, and neglecting the mass of the electron and nucleon, the corresponding differential cross section can
be written
\bea
2\omega_{k'}\, \frac{d\sigma}{d^3k'}  
=\ \frac{1}{2^2}\frac{1}{2s}\frac{1}{(2\pi)^3 }\
\sum_X\sum_{\lambda,\lambda',\sigma}\ \left | A_{eN\rightarrow eX} \right|^2\; (2\pi)^4\, \delta^4(p_X+k'-p-k)\, ,
\label{eq:e-diff}
\eea
with $A_{eN\to eX}$ given by Eq.\ (\ref{eq:ampl}).  In this cross section we use what we know from QED to separate a leptonic tensor, which we easily calculate, from the ``unknown" hadronic tensor, 
\bea
2\omega_{k'} \frac{d\sigma}{d^3k'} = \frac{1}{s(q^2)^2}\ L^{\mu\nu}W_{\mu\nu}\, .
\label{eq:e-LW}
\eea
To be specific, the leptonic tensor is
\bea
L^{\mu\nu} =  \frac{e^2}{8\pi^2}\sum_{\lambda,\lambda'}
[\bar{u}(k',\lambda')\gamma^\mu u(k,\lambda)]^*\, \bar{u}(k',\lambda')\gamma^\nu u(k,\lambda)
= 
 \frac{e^2}{2\pi^2}\, \left(\, k^\mu k'\, {}^\nu + k'\, {}^\mu k^\nu - g^{\mu\nu}k\cdot k'\, \right)\, .
 \label{eq:lep-tensor}
\eea
This leaves us with the star of the show, the hadronic tensor, written in terms of the unknown matrix elements of the current,
\bea
W_{\mu\nu}^{N} &=& \frac{1}{8\pi}\ \sum_{\sigma,X}\ \langle X|J_\mu(0) |p,\sigma\rangle^*
\langle X|J_\nu(0) |p,\sigma\rangle
\ \int d^4y\, e^{-i(p_X-p-q)\cdot y}
\nn\\[2mm]
&=& \frac{1}{8\pi}\ \int d^4y\, e^{iq\cdot y} \sum_{\sigma,X}\ \langle p,\sigma|J_\mu(y) | X\rangle
\langle X|J_\nu(0) |p,\sigma\rangle
\nn\\[2mm]
&=& 
\frac{1}{8\pi}\ \sum_\sigma \int d^4 y \, e^{iq\cdot y}\ \langle p,\sigma |\, J_\mu(y) \, J_\nu(0) |p,\sigma\rangle\, ,
\label{eq:W-def}
\eea
where in the first equality we expand the momentum-conservation
delta function as a coordinate-space integral, using $k'-k=-q$.
In the  second expression we have  used the hermiticity of the current operator and translation invariance,
and in third expression we have summed over final states, $X$.
 We have dropped the label ``EM" from the current, since the same reasoning applies to the currents
 mediating the weak interactions.
 
 The hadronic tensor defined in this way is dimensionless.  Although it is a four-by-four matrix, the
 symmetries of the strong interactions and conservation of the electromagnetic current imply that
 it can be written in terms of a limited number of tensors, each multiplied by a scalar ``structure function",
 \bea
W^{Vh}_{\mu\nu} &=& -\left( g_{\mu\nu} - \frac{q_\mu q_\nu}{q^2}\right)\, F_1^{Vh}(x,Q^2)
+ \left(p_\mu-  q_\mu\, \frac{p\cdot q}{q^2}\right)\,
\left(p_\nu-  q_\nu\, \frac{p\cdot q}{q^2}\right)\, \frac{F_2^{Vh}(x,Q^2)}{p\cdot q}
\nn\\[2mm]
&\ & \quad
-\ i\epsilon_{\mu\nu\lambda\sigma}p^\lambda q^\sigma\ \frac{F_3^{Vh}(x,Q^2)}{p\cdot q}\, 
+ \epsilon_{\mu\nu\lambda\sigma} q^\lambda \, \left ( \frac{s^\sigma}{p\cdot q} g_1(x,Q^2) +\frac{ \left [ p\cdot q s^\sigma - s\cdot q p^\sigma \right ]}{ (p\cdot q)^2 } g_2(x,Q^2) \right )
\nn \\
\label{eq:F123}
\eea
where the $F_i$, which are also dimensionless, are chosen to depend on $Q^2$
and the dimensionless variable $x=Q^2/2p\cdot q$.  This ratio, the
``Bjorken scaling variable" \cite{Bjorken:1968dy}, will appear naturally below. The function $F_3$ contributes only through the parity-non-conserving
weak interactions.  The final two terms are present for targets with spin.   We also show two (parity conserving) structure functions, $g_1$
and $g_2$ that require hadronic polarization in the initial state; for spin-dependent parity-violating structure functions, see Ref.\ \cite{Anselmino:1993tc}.

\subsection{Scaling and the light cone}

Until the experiments were actually performed, no one knew quite what to expect, but the kinematics of the process can be suggestive.
In DIS, $q^2\equiv -Q^2$ is negative.  We can always go to a frame where $p^\mu$
is in the plus direction and $q_\perp=0$.
In such a frame, the momentum transfer $q^\mu$ has two light-cone components, plus and minus\footnote{Our light cone 
coordinates are $q^\pm = (1/\sqrt{2})(q^0\pm q^3)$, with $q^2=2q^+q^--q_T^2$.}.   The minus component
must be positive
 in such a frame, so that the final state momentum $p_X$, which is massive, has positive plus and minus momenta.
This requires $q^+$ to be negative, and we can always write $q^+=- xp^+$, for some $0 < x <1$.  So far, this is just notation in the particular frame.

We now note, however, that we can express the variable $x$ as
\bea
x = - \frac{q^+}{p^+} = \frac{Q^2}{2p\cdot q}
\label{eq:x-def}
\eea
and that $(xp+q)^2=0$.  This identity can be given a physical interpretation in the matrix elements of
Eq.\ (\ref{eq:W-def}), as follows.  The electromagnetic current on the right, $J_\nu(0)$, can be pictured as absorbing a fermion at 
one momentum, then reemitting it with that momentum plus $q$.   It is then natural to picture the 
action of the currents in (\ref{eq:W-def}) as the absorption of a particle -- a {\it parton} -- with fraction $x$ of the nucleon's
momentum $p$, and the emission of the same particle, now with momentum $xp+q$.   The 
current on the left, $J_\mu(y)$ absorbs the scattered particle, transferring its momentum back to $xp$,
and reforming the nucleon.   If the particle is to propagate between the two currents, they should be
separated by a light-like distance, $y^\mu,\, y^2=0$, which should be in the {\it minus} direction, opposite
to the direction of the nucleon.   Then $y^+\sim 0$, and the factor $e^{iq^-y^+}$ is unity.  In this scenario,
the Fourier transform of the hadronic tensor is independent of $q^-$, and therefore of the actual value
of $Q^2$.   All of its $q$-dependence is in terms of the variable $x$, the fractional momentum
of our scattered particle.   It's important to emphasize that the usefulness of this picture is not obvious,
and that in fact early experimental results and theoretical expectations  developed simultaneously and
with constructive interplay \cite{Bjorken:1968dy,Panofsky:1968ka}.   Only later, with the discovery of asymptotic
freedom \cite{Gross:1973id,Gross:1973ju,Gross:1973zrg,Politzer:1973fx,Georgi:1976ve}, could calculation proceed from first principles.

A persuasive way to motivate the role of light-like separations between the currents in Eq.\ (\ref{eq:W-def}) is to recall
that the coordinate-space propagators of field theory are themselves singular on the light cone.  So, for
example, the Fourier transform of $\rlap{p}{/}/(p^2-m^2+i\epsilon)$ to coordinate $y$ behaves like $\rlap{y}/ /(y^2)^2$
for small $y^2$, even when the mass $m$ is nonzero.  Perturbative considerations like this suggest
that bilocal matrix elements like (\ref{eq:W-def}) are dominated by light-cone separations.    Schematically, if
the electromagnetic current is found from a sum over partonic fields $\psi_a$, we can use a ``light-cone expansion" 
\cite{Leutwyler:1970vr,Frishman:1970fx,Brandt:1970kg,Christ:1972ms}, of
the general form,
 \bea
 \langle p,\sigma |\, J_\mu(y) \, J_\nu(0) |p,\sigma\rangle_{{y^+,y_\perp \to 0}} \
{ \rightarrow}\ \sum_a C^{(a)}_{\mu\nu}(y^2,y\cdot p)\  \langle p,\sigma |\, \psi_a^\dagger (y^-)\, \rlap{y}/ \,  \psi_a(0) |p,\sigma\rangle\, ,
\label{eq:light-cone}
 \eea
 where the Dirac matrix $\rlap{y}/ $ can be thought of as the numerator of the propagator of the scattered quark, and
 where the tensor $C_{\mu\nu}(y^2,y\cdot p)$ carries the singular behavior as $y^2\to 0$.  This is certainly the case in a free field theory.
 One might ask, why we should care about a free field theory?   Experiment required us to do so! \cite{Bloom:1969kc,Breidenbach:1969kd}
 
 As suggested above, the  DIS data from SLAC exhibited the property of scaling,
 the (relatively accurate) independence of $Q^2$ for the measured structure functions $F_1$ and $F_2$ .
 It was as if the strongly-interacting constituents of the nucleon were scattering from
 electrons like free particles, on the face of it quite a paradoxical result.   
Nevertheless, the physical  picture of DIS on any hadron $h$ given above is summarized by a formula for
electroweak scattering in the {\it parton model},
\bea
\sigma^{\rm incl}_{\rm eh}(p,q)\ =\ 
\sum_{{\rm partons}\ a} \int_0^1 d\xi\, \hat{\sigma}^{\rm el}_{ea}(\xi p,q)\; \phi_{a/h}(\xi)\, ,
\label{eq:parton-sigma}
\eea
where $\sigma^{\rm incl}_{eh}(p,q)$ is the inclusive cross section for 
$e(k) + h(p) \rightarrow e(k'=k-q) + X(p+q)$ with DIS kinematics, while
$\hat \sigma^{\rm el}_{ea}(\xi p,q)$ is the elastic cross section
$e(k) + a(\xi p) \rightarrow e(k'-q) + a(\xi p+q)$ which sets 
$(\xi p+q)^2=0 \rightarrow \xi=-q^2/2p\cdot q = x$.  

For the unpolarized cross section with a measured electron, Eq.\ (\ref{eq:e-LW}), for photon exchange,
we can compute $W_{\mu\nu}^{\gamma f}$, directly in the parton model, and determine the parton model structure functions, 
$F_i$ from the relation
\bea
W_{\mu\nu}^{\gamma h}(p,q)\ =\ \sum_{{\rm partons}\ a} \int _0^1\frac{d\xi}{\xi}\, W_{\mu\nu}^{\gamma a}(\xi p,q)\, \phi_{a/h}(\xi)\, .
\eea
The extra factor of $1/\xi$ reflects the difference between the invariant normalization
of the hadronic cross section, $1/2s$ in Eq.\ (\ref{eq:e-diff}), and the corresponding 
normalization for the partonic cross section, $1/2\xi s$.
In these relations for the structure functions and cross sections, $\phi_{a/h}(\xi)$ 
has the interpretation of the  ``probability density for a parton of type $a$
 to have momentum $\xi p$ of hadron $h$".   It is  independent of the details of the hard scattering.
This is  the hallmark of factorization.   In the parton model, the strong interactions
determine the functions $\phi_{a/h}(\xi)$, but play no role in the electroweak
scattering $\hat\sigma^{\rm el}_{ea}$.

We can compute the structure functions directly from the tree-level hadronic tensor,
\bea
W_{\mu\nu}^{\gamma a}(\xi p,q)
&=& \frac{1}{8\pi}\ \sum_\sigma \int d^4 y \, e^{iq\cdot y}\ \langle \xi p,\sigma |\, J_\mu(y) \, J_\nu(0) |\xi p,\sigma\rangle^{(0)}
\nn\\[2mm]
&=&
\frac{e_f^2}{8\pi}\ \sum_{\sigma,\sigma'} \int \frac{d^4p'}{(2\pi)^4}\, (2\pi)\delta(p'{}^2) 
\langle \xi p,\sigma |\, \bar \psi_f \gamma_\mu \psi |p',\sigma'\rangle^{(0)} \langle p',\sigma'| \, \bar \psi_f \gamma_\nu \psi |\xi p,\sigma\rangle^{(0)}
\nn\\
&\ & \hspace{40mm} \times\ (2\pi)^4\delta^4(p'-\xi p-q)
\nn\\[2mm]
&=& -\left( g_{\mu\nu} - \frac{q_\mu q_\nu}{q^2}\right)\, \delta\left (1-\frac{x}{\xi}\right) \frac{e_f^2}{2}
\nonumber\\
&\ & + \left(\xi p_\mu-  q_\mu\, \frac{\xi p\cdot q}{q^2}\right)\,
\left(\xi p_\nu-  q_\nu\, \frac{\xi p\cdot q}{q^2}\right)\, \delta\left ( 1-\frac{x}{\xi}\right) 
\frac{e_f^2}{\xi p\cdot q}\, ,
\label{eq:quark-W}
\eea
with, as above, $x=Q^2/2p\cdot q$ and $q^2 =-Q^2$, and where the hadronic momentum appearing in the tensors is $\xi p$.
The left-over delta functions are a direct result of fixing the momentum of the outgoing electron.

With $W^{\gamma a}$ in hand, we can read off the parton model structure functions from  Eq.\ (\ref{eq:F123}), keeping in mind that the parton's momentum is $\xi p$,
\bea
F^{\gamma h}_2(x)\ =\ 2xF^{\gamma h}_1(x)\ =\ \sum_{{\rm quarks} f} e_f^2 x\, \phi_{f/h}(x)\, .
\label{eq:C-G-relation}
\eea
The delta functions in the hadronic tensor Eq.\ (\ref{eq:quark-W}), which 
set $\xi=x$, eliminate the integrals.
As expected, the parton model structure functions are 
independent of $Q^2$,
the property called scaling.

The proportionality between $F_1$ and $F_2$ for our spin-1/2 quarks in Eq.\ (\ref{eq:C-G-relation})
is a key result, the Callan-Gross relation \cite{Callan:1969uq}.
 This relation is quite different for scalar quarks.
 The Callan-Gross relation shows the compatibility 
of the quark and parton models.

 In summary, in the parton model, each hadronic deep inelastic cross section is a
sum of convolutions of partonic elastic cross sections 
with the hadron's parton distributions.
 The very nontrivial assertion is that there is a quantum mechanical  incoherence  between
large-$q$ scattering and the partonic distributions.
In effect, we are multiplying probabilities rather than adding amplitudes and then squaring.
Its heuristic justification is that  the binding of the nucleon involves
 long-time processes that do not interfere with the 
 short-distance
 scattering.   We imagine that, in its rest frame, the nucleon consists of a superposition of partonic
 states that have some maximum virtuality, $\Delta E_{\mathrm{max}}$.  Such
 states then have lifetimes down to $1/\Delta E_{\mathrm{max}}$ in the rest frame.   But in the
 rest frame of the electron in DIS, or even in the center of mass frame, 
 such states have lifetimes that are time dilated.   When $1/Q$ is much
 smaller than all dilated times, the exchange of a virtual photon will take
 place when the nucleon is ``frozen" in one of those states.   This of course
 is a simplified picture, but as we shall see it can serve as a starting point
 for the treatment in quantum field theory.
 
 \section{From partons to QCD}
 \label{sec:partons-to-qcd}
 
 \subsection{The reconciliation of scaling and strong interactions}

 The paradox of the successes of the parton model, with its free scattering for strongly-interacting 
 particles, was resolved not long after by the discovery that nonabelian gauge
 theories can be asymptotically free, so that at the short distances and times
 during which the high-$Q^2$ virtual photon of DIS is absorbed, charged
 quarks (as the partons turned out to be) really do act as if they were free, up to
 calculable corrections.   The systematics of these corrections are what
 we now term collinear factorization \cite{Collins:1989gx,Sterman:2022gyf} and its (DGLAP) evolution \cite{Altarelli:1977zs,Gribov:1972ri,Dokshitzer:1977sg}. 
 
 For the purposes of this discussion, we only recall that the perturbative strong coupling in QCD decreases 
 logarithmically as the energy scale at which it is defined increases, in the lowest approximation as,
 \bea
 \as(\mu_2)\ =\ \frac{\as(\mu_1)}{1+b_0\, \as(\mu_1)\, \ln\frac{\mu_2^2}{\mu_1^2}}\ =\ 
 \frac{1}{b_0\ln \left(\frac{\mu_2}{\Lambda_{\tiny \rm QCD}}\right)^2 }\, ,
 \label{eq:as-1loop}
 \eea
 with $b_0=(33-2n_F)/12\pi$ for QCD with $n_F$ quark flavors with masses below $\mu_2$.
  Correspondingly, the coupling increases
 for lower momentum scales, diverging altogether at the scale denoted $\Lambda_{\mathrm{QCD}}$,
 of the order of 200 or 300 MeV, depending on the number of relevant quark flavors.  (Note that $\Lambda_{\tiny\rm QCD}$ is
 independent of $\mu_1$.)   We can
 see the utility of this property by considering a cross section or other physical
 quantity that depends on a single kinematic scale $Q$, computed in perturbation theory.
 The general form of such a quantity is
 \bea
\sigma\left(\frac{Q^2}{\mu_R^2},\frac{m_i^2}{\mu_R^2},\alpha_s(\mu_R),Q\right)
&=&
\sum_{n=1}^\infty a_n\left(\frac{Q^2}{\mu^2_R}, \frac{m_i^2}{\mu_R^2},Q\right)\, \alpha_s^n(\mu_R)\, ,
\label{eq:sigma-expand}
\eea
where $m_R$ is the renormalization scale and $m_i$ denotes the fixed masses in the problem, including in principle the zero gluon
mass of QCD.  A fundamental result of quantum field theory is that a physical quantity
cannot depend on our choice of the renormalization scale,
\bea
\frac{ d\sigma}{d\mu_R}\ = 0\, .
\label{eq:mu-independence}
\eea
Given our freedom in choosing $\mu_R$, we'd like to choose it to make the coupling
$\as(\mu_R)$ as small as possible, so that our series can converge as fast as possible, but generally not larger than 
the scale $Q$, to avoid logarithms of the ratio $Q/\mu_R$.
If we choose $\mu_R$ at scale $Q$, however, the ratios $Q/m_i$ will be large, leading to large coefficients $a_n$.
The art of using asymptotic freedom is to identify quantities which do not depend logarithmically on fixed mass scales, \cite{Sterman:1975xv,Sterman:1977wj}
\bea
a_n\left(\frac{Q}{\mu_R}, \frac{m_i^2}{\mu_R^2},Q\right)\, \alpha_s^n(\mu_R)
=
a_n\left(\frac{Q}{\mu_R}, 0,Q \right)\, \alpha_s^n(\mu_R)\ +\ 
{\cal O}\left( \left[\frac{m_i^2}{\mu_R^2}\right]^p\right)\, ,
\label{eq:an-af}
\eea
for some positive power $p$.   Such quantities are conventionally said to be {\it infrared safe}.   The parton 
model showed the way to identify families of such quantities, which generalize the cross sections
of free partons.

\subsection{Factorization generalizes the parton model}
 
 For DIS, it's convenient to phrase factorization in terms of the structure functions, and
 for each $F_i$ of a photon on hadron $h$, we have
 \bea
F_i^{\gamma h}(x,Q^2) &=& 
\sum_{\mathrm{partons}\ a}\int_x^1 \frac{d\xi}{\xi^{1-\delta_{i2}}} \; C_i^{\gamma a}
\left( {x\over \xi},{Q\over \mu_R},{\mu_F\over \mu_R},\alpha_s(\mu_R)\right)
 \ \phi_{a/h}(\xi,\mu_F,\alpha_s(\mu_R))
\nonumber\\[2mm]
&\equiv& C_i^{\gamma a}
\left( {x\over \xi},{Q\over \mu_R},{\mu_F\over \mu_R},\alpha_s(\mu_R)\right)
\otimes \phi_{a/h}(\xi,\mu_F,\alpha_s(\mu_R))\, ,
\label{eq:F2-fact}
\end{eqnarray}
 where the second line is a common notation for convolutions in momentum fraction $\xi$.\footnote{The 
 lack of a $1/\xi$ for the $F_2$ convolution reflects the overall linear power of hadronic momentum in the coefficient of $F_2$ in 
 Eq.\ (\ref{eq:F123}).}
Such a convolution form in fractional momenta is often termed ``collinear factorization".  Compared to 
 the parton model template, Eq.\ (\ref{eq:parton-sigma}), we have introduced dependence on the strong coupling, $\as(\mu_R)$,
 and a ``factorization scale", $\mu_F$.   Roughly speaking, we want to separate dynamics at mass scales $\mu_F$ and above
 into the ``coefficient" or ``short  distance", functions $C_i^{\gamma a}$, which are infrared safe and so can be calculated in perturbation theory.
 As in the parton model, the $\phi_{a/h}$ cannot be computed directly in perturbation theory, although, as we shall
 see, their dependence on $\mu_F$ is amenable to computation.  In this expression, it will be natural to pick both the
 renormalization and factorization scales of the same order as the single hard scale, the momentum transfer $Q$.
 To simplify the discussion, let's pick $\mu_R=\mu_F$, and concentrate on the resulting $\mu_F$ dependence.   Our goal below is
 to see the nature of $\mu_F$ dependence in DIS, which will provide a template for the use of perturbative QCD
 in other hard-scattering processes.   For this purpose, we need to specify what we 
really mean by factorization, introducing the concept of a scheme as well as a  scale.

\subsection{Cycling twice through the factorization formulas}

To anticipate, we are going to cycle  the factorized cross section, Eq.\ (\ref{eq:F2-fact}), twice.   The first cycle will be in an imaginary world,
with an ``infrared-regulated" version of QCD, in which we can calculate both the parton distributions $\phi_{a/b}$ for
 distributions of partons in partons and $F_i$ itself for parton scattering.
This will require us to define what we mean by our parton distributions, and this
definition defines the factorization scheme.  In this world, and in this scheme, we compute the coefficient 
functions $C_i$, and confirm that they are infrared safe.   This concludes the first cycle.\

Once the first cycle is complete, we use the infrared safe $C_i$ in the same factorized structure function
Eq.\ (\ref{eq:F2-fact}) again, this time in the real world.  Because the $C_i$ are IR safe, we assume that they are the same
in the IR regulated and real theories.
We can't calculate either $F_2$ or the parton distributions in our real world, but we can {\it measure} the
structure functions (if we have an appropriate accelerator at hand!).   Once we've done that, we can determine
the parton distributions by combining those measurements
with the calculated coefficient functions.   
In recent years, ``measurement" can include lattice simulations \cite{Ji:2013dva,Radyushkin:2017cyf,Lin:2017snn,Briceno:2017max}.
As we shall see, the distributions so determined are ``portable" to a large
set of other processes {\it and} other energies.

This ``double cycle" uses calculation and measurement hand-in-hand to determine the parton distributions of
the real world.   It is the template for all applications of factorized cross sections, and the foundation for
precision measurements using other electroweak interactions with hadrons. 

\subsection{$F_2$ at zeroth and first orders}

For this purpose, we will describe how we compute the first correction to $F_2^{\gamma q_f}(x,Q)$ in the infrared-regulated theory.  That is, we compute the
structure function for the particular hadron $h=q_f$, one of
the quarks, say flavor $f$ (a light one, because we will neglect its mass).   Let's see how this works for our quark, absorbing  a space-like photon.  

Our general factorized structure function, Eq.\ (\ref{eq:F2-fact}) can now be expanded in $\as$ as,
\begin{eqnarray}
\hspace{-10mm}
F_2^{\gamma q_f}(x,Q^2) &=& C_2^{(0)}\, \otimes\, \phi^{(0)}
+ {\alpha_s\over 2\pi}\ 
C_2^{(1)}\, \otimes\, \phi^{(0)}
+ {\alpha_s\over 2\pi}\ C_2^{(0)}\, \otimes\, \phi^{(1)} + \dots\, ,
\label{eq:F2-expand}
\end{eqnarray}
and so on.
At  zeroth order in QCD, there are no strong interactions, and we
get just the parton model result, in which the coefficient function is simply a charge factor times a delta function.
By comparison to the parton model tensor, Eq.\ (\ref{eq:quark-W}), and postulating that the 
probability distribution of a free  parton ``in itself" is a delta function, 
\bea
\phi_{q_f/q_{f'}}^{(0)}(\xi)=\delta_{ff'}\; \delta(1-\xi)\, ,
\label{eq:phi-zero}
\eea
 we can read off the zeroth order, parton model, coefficient functions.
\bea
C^{\gamma q_f(0)}_2\left (\frac{x}{\xi}  \right)\ =\ e^2_f\, \delta \left (1-\frac{x}{\xi}  \right)\ =\ 2 C_1^{\gamma q_f(0)} \left (\frac{x}{\xi}  \right)
\, .
\label{eq:C-zero}
\eea
 Substituted into Eq.\ (\ref{eq:F2-fact}), the zeroth order coefficient functions of perturbative 
 QCD reproduce the parton model structure functions in
  Eq.\ (\ref{eq:C-G-relation}).

To go beyond lowest order in the strong interactions in an infrared-regulated
version of QCD (first ``cycle" above), we use the approach that has been extensively applied to 
electroweak and strong scattering processes at all high-energy colliders,
relying on QCD in $D$ dimensions.
For the resulting perturbative amplitudes we can prove factorization theorems, so we
know what to look for, in particular infrared safe coefficient functions, like the one in Eq.\ (\ref{eq:F2-fact})
for the $F_i$ in DIS.  Once we have isolated these quantities, we apply them in the
same factorization theorems with actual data, to determine physical parton distributions for real hadrons (second ``cycle").   
It is important to emphasize that factorization proofs are carried out for partons in
the regulated theory, and we are assuming that the same infrared safe functions
apply for hadronic as for partonic scattering.  

\begin{figure}
\centerline{\figscale{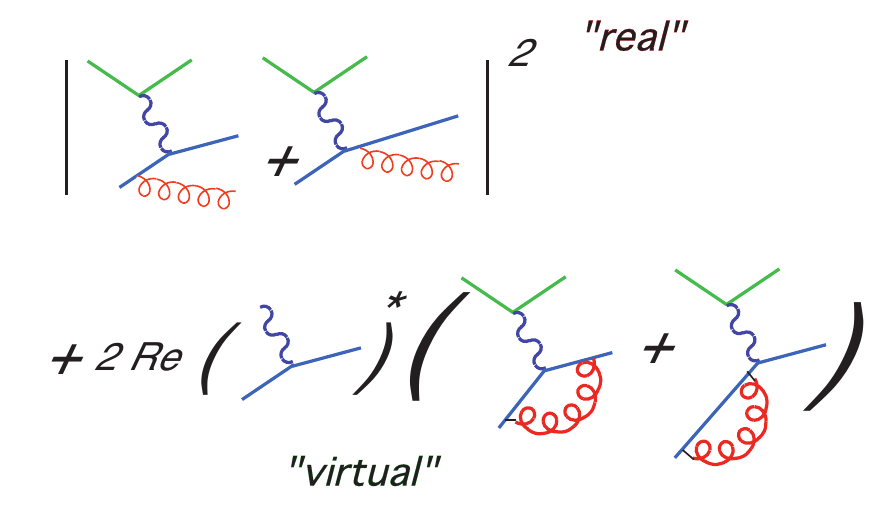}{10cm}}
\caption{Order $\as$ diagrams for the hadronic tensor of DIS. \label{fig:order-as-dis}}
\end{figure}

With all this in mind, we go on to order $\alpha_s$, in Eq.\ (\ref{eq:F2-expand}) 
for $F_2^{\gamma q_f}$, where we will encounter the choice of scheme.
The diagrams that contribute to the hadronic tensor and hence structure functions 
at this level are shown in Fig.\ \ref{fig:order-as-dis}.
Because our cross section is inclusive, we have to combine final states with different phase space, both of
which, as we shall see, are formally divergent in four dimensions.
Recalling the lesson from QED -- inclusive cross sections summed to all orders
can reduce approximately to elastic scattering -- we anticipate at least a partial cancellation
of these divergences.   A convenient notation that organizes such cancelations
is provided by the generalized functions known as `plus distributions", defined by their integrals with smooth functions,
for example with any integer $n$,
\bea
\int_0^1 dx\ f(x){\left(\ln^n(1-x)\over 1-x\right)_+} \ &\equiv& \
\int_0^1 dx\ \left(\, f(x)-f(1)\, \right)\ {\ln^n(1-x) \over (1-x)}\, .
\label{eq:plus-defs}
\eea
For us, in deep inelastic scattering, $f(x)$ will be a parton distribution.
The term proportional to $f(x)$ will correspond to a gluon emitted into the final state, with momentum fraction 
$1-x$, while the $f(1)$ term corresponds to a virtual correction, with elastic kinematics in the cross section.
An important feature of the plus distributions is that their integrals with constant functions vanish.
But for us,  $f(x)$ is a parton distribution, and hence not a constant.

Combining the contribution from the diagrams in Fig.\ \ref{fig:order-as-dis} to the structure function $F_2$  with the
lowest order gives,
\begin{eqnarray}
F_2^{\gamma q_f}(x,Q^2) &=&  e^2_f\ \Bigg\{\ x\ \delta(1-x)
+ {\alpha_s\over 2\pi}\
 C_F\Big ( (1+x^2) \left[ {\ln(1-x)\over 1-x} \right ]_+ -\frac{3}{2}\left[\frac{1}{1-x}\right]_+ 
 \nonumber\\ 
&\ &
\nonumber\\ 
&\ & \hspace{-20mm}  - \left( \frac{9}{2}+\frac{\pi^2}{3} \right) \delta(1-x) +{\cal C}(x)
  \Big )
 + \
{\alpha_s \mu_F^{4-D}\over 2\pi^2 }\ 
 C_F\, \left [\int_{\mu_F}^{Q} + \int_{0}^{\mu_F} \right ]{d^{D-2}k_T^2\over k_T^2}\
\left[{1+x^2\over 1-x}\right]_+\ \Bigg\}\, ,
\nn\\
\label{eq:F2-oneloop}
\end{eqnarray}
where ${\cal C}(x)$ is a smooth function.   The constant $C_F=(N_c^2-1)/2N_c=4/3$
with $N_c=3$ colors for QCD.
The first term on the right-hand side is the zeroth order.   Following this, we
have a set of finite terms, which we can group in the coefficient function $C_2^{(1)}$ of
Eq.\ (\ref{eq:F2-expand}).   We have split the final term, which is proportional to
an integral over the gluon's transverse momentum that diverges logarithmically in four dimensions,
into two parts by introducing a scale $\mu_F$, which we identify as the factorization
scale.   This separation constitutes a ``scheme" to define the parton distribution,
and hence the coefficient function, here to order $\alpha_s$.   The separation
is by no means unambiguous -- we can always shift finite terms between the coefficient
function and the parton distribution.   Here, we are choosing a  ``minimal" prescription,
for which (we keep only the pole in $\epsilon\equiv 2-D/2$ in)
\begin{eqnarray}
\phi_{q/q}^{(1)}(x,\mu^2) = {\alpha_s \mu_F^{4-D}\over \ 2\pi^2}\ P^{(1)}_{qq}(x)\, \int_0^{\mu_F^2}\,
{d^{D-2}k_T^2\over k^2_T}\, ,
\label{eq:msbar-phi1}
\end{eqnarray}
which is proportional to the ``DGLAP evolution kernel" or ``splitting function",
\begin{eqnarray}
P_{qq}^{(1)}(x) =  C_F\, \left( \left[{1+x^2\over 1-x}\right]_+ +\ \frac{3}{2}\; \delta(1-x) \right )\, .
\label{eq:Pqq}
\end{eqnarray}
The corresponding ``minimal" coefficient function is then defined in four dimensions as
 \bea
 C_2^{(1)}(x) &=& (\alpha_s/2\pi)\ P_{qq}(x) \ln(Q^2/\mu_F^2)
 \nn\\
 &\ & \hspace{-10mm}
 + {\alpha_s\over 2\pi}\
 C_F\Big ( (1+x^2) \left[ {\ln(1-x)\over 1-x} \right ]_+ -\frac{3}{2}\left[\frac{1}{1-x}\right]_+ 
  - \left( \frac{9}{2}+\frac{\pi^2}{3} \right) \delta(1-x) +{\cal C}(x)
  \Big )\, .
  \label{eq:C2-1}
\eea
The IR safety of the coefficient function relies on the cancellation of gluon
emission in the $x\to 1$ limit.   This is the limit in which the emitted gluon
is ``soft", carrying negligible energy.   This cancellation is the QCD analog
of the cancellation of soft photons in QED.   The inclusivity of the cross section
has had the desired effect.   At the same time, a singlular $k_T$ integral
remains at finite $x$, corresponding to the emission of a finite-energy
photon parallel, or ``collinear" to the incoming quark at arbitrarily early times.  These initial-state
collinear singularities do not cancel in the sum over states, and are the
practical reason we need to factorize the cross section.  We will come
back to the cancellation of soft singularities and their separation from
collinear singularities beyond lowest order in the following subsection.

Finally, we note that at one loop $F_1$ is given in terms of $F_2$ by
\begin{eqnarray}
F_1^{\gamma q_f}(x,Q^2) &=&
{1\over 2x}\ \left\{ F_2^{\gamma q_f}(x,Q^2)\right \} - 
e_f^2\, C_F\, {\alpha_s\over \ 2\pi}\; x\, .
\eea
The  same minimal factorization scheme 
and parton distributions can be used for both structure functions, and the
Callan-Gross relation, Eq.\ (\ref{eq:C-G-relation}), holds up to
a calculable correction in QCD.

The advantage of the minimal parton distribution is its relative simplicity,
and its independence of the process -- it is linked to the DIS cross section
only through the scaleless variable $x$ and the strong coupling.   In fact,
the one-loop expression for $\phi_{q_f/q_f}$ is the order $\as$ expansion of a matrix element
that recalls the ``light cone" form of the free theory, Eq.\ (\ref{eq:light-cone}).
Quite generally, we define the quark and gluon distributions as matrix elements \cite{Collins:1981uw},
\bea
\phi_{a/h}(x,\mu_F)
&=&
\frac{1}{2}\, \sum_{{\rm spins}\ \sigma}\
\int \frac{dy^-}{2\pi}\, e^{-ixp^+y^-}\ 
\langle p,\sigma | \bar q_a(y^-)\ Pe^{-ig\int_0^{y^-}dy^- n\cdot A^{(F)}(y^-n)}
\frac{\gamma^+}{2}\, q_a(0)| p,\sigma\rangle\, ,
\nn\\
\phi_{g/h}(x,\mu_F)\ 
&=&\ \int \frac{dy^-}{2\pi}\, e^{-ixp^+y^-}\ 
\langle p,\sigma | \bar F^+_\mu(y^-)\ Pe^{-ig\int_0^{y^-}dy^- n\cdot A^{(A)}(y^-n)}
\, F^{\mu +}(0)| p,\sigma\rangle\, ,
\label{eq:phi-q-msbar}
\eea
where $n^\mu$ is a unit light-like vector in the direction opposite to $p^\mu$.
At zeroth order these distributions reduce to the delta function, Eq.\ (\ref{eq:phi-zero}),
and beyond lowest order they can be defined by a generalization of the minimal prescription above,
with $\mu_F$ playing the role of a renormalization scale.
The path-ordered exponential of the gauge field that connects the two
parton fields, $q(0)$ and $\bar q(y^-)$ in (\ref{eq:phi-q-msbar}) is in the fundamental (quark) representation, and is
a consequence of the gauge
nature of QCD.  
Similarly, the path ordered exponential in the gluon distribution is in adjoint representation.
 In QCD, computations in perturbation theory require
picking a gauge.  In gauges that maintain Lorentz invariance,
on-shell massles charged particles are accompanied
by collinear gluons with unphysical polarizations, proportional to their momentum.   In the
high energy, or massless, limit, these gluons are associated with collinear singularities.
It is precisely because they are by themselves unphysical that their effects are summarized
by the ordered exponential moving in the direction opposite to the physical charged particle.
The incoming particle couples to the external world as if the world were a point source of
color charge moving in the opposite direction. 

The determination of coefficient functions at a given loop order completes a 
first ``cycle" of the analysis of the factorized cross section summarized by
Eq.\ (\ref{eq:F2-fact}), here at order $\as$.   
For DIS, coefficient functions are known to order $\as^3$ \cite{Vermaseren:2005qc,Blumlein:2022gpp}.
In the second cycle, one begins with the order $\as$
coefficient functions, {\it measures} the structure functions, and from
that determines the parton distributions.   Once coefficient functions to a given order are
determined in DIS and/or a set of other measured processes, systematic fits to the distributions are possible
\cite{Accardi:2016ndt,Kovarik:2019xvh}.

\section{Factorization and evolution}
\label{sec:fact-to-evol}

We've now seen how factorization can work at the first nontrivial order in $\as$, and
we'd like to argue that this pattern extends to arbitrary order.   The essential
observation is that in quantum field theory there is no ``shortest-lived" state 
that makes up a nucleon.   Rather, a nucleon may arrive at the scattering
in a state of any virtuality.   Indeed, for scatterings of increasing 
momentum transfer, requiring shorter and shorter distances and times, and hence higher virtuality,
the numbers of accessible high energy states grows.  

Looking at the first QCD correction to DIS in Fig.\ \ref{fig:order-as-dis}, we see the
basic process as first the emission of a gluon from an on-shell quark, followed by
the absorption of the photon, which ``frees" the gluon.   As $Q^2$ increases, the gluon
so freed can have larger and larger transverse momentum relative to the 
parent quark, and hence emerge from virtual states of shorter and shorter mean
lifetimes.  But as $Q^2$ increases the number of virtual states also increases,
which overcomes the suppression associated with their virtuality.  There is, in
the discussion at the end of Sec.\ \ref{sec:j-to-partons}, no $\Delta E_{\mathrm{max}}$, or
largest virtuality among states that mix with quarks, and hence with the nucleon.
In quantum field theory, the whole universe of particles flickers in and out of existence, and
we can detect them if only we can search at time scales short enough.

Of course, at any fixed momentum transfer, longer-lived states still contribute, corresponding
to low values of gluon transverse momentum in Fig.\ \ref{fig:order-as-dis}.   As illustrated in
Eq.\ (\ref{eq:F2-oneloop}), as this transverse momentum vanishes, it produces collinear enhancements,
which we organize into parton distributions in the regulated theory, enabling us to 
calculate the dependence of coefficient functions on the hard scale.

At this point we'll give a sense of how to systematize all orders in perturbation theory,
a taste of one approach to collinear factorization proofs in perturbative QCD and other field theories.
The arguments apply directly to the infrared regulated version of the theory,
given entirely in perturbative terms.  Our goal is to argue that all infrared
dependence of the DIS structure functions (and hence cross section) is absorbed into the parton distributions
of the factorized expression, Eq.\ (\ref{eq:F2-fact}).   

\subsection{Factorization beyond lowest order}

To give an all-order argument, we need an understanding of the origin of infrared singularities in QCD.
A full analysis, which depends on the behavior of integrals in multidimensional
complex space, is not possible here.  Introductions to the approach can be found in \cite{Collins:1989gx} and \cite{Sterman:1995fz},
and a full development in \cite{Collins:2011zzd}.
In fact, we can summarize the result of these considerations
in a simple rule.   The rule applies both to (Feynman) diagrams for amplitudes and to ``cut diagrams",
which combine amplitudes with complex conjugates.   
For any (cut) diagram, as we integrate over loop and phase space variables, the
only sources of infrared divergences are momentum configurations that correspond to free,
classical, on-shell propagation between vertices.  
At these points in momentum space, we can actually assign coordinate space positions
to vertices, and each particle travels with a definite momentum at its classical velocity between these points.
This result, identified in this
form by Coleman and Norton \cite{Coleman:1965xm}, applies to both the massive and massless
lines, but is particularly powerful for massless particles.  In applications to cross sections, particles in the amplitude flow
forward in time between vertices, while in the complex conjugate amplitude they 
flow backward.

A schematic cut diagram for DIS in electron-quark scattering is shown in Fig.\ \ref{fig:dis-fact}a.  The amplitude ($A$) and
complex conjugate ($A^*$) are separated by a vertical line that ``cuts" the diagram and
identifies the final state to which it contributes \cite{Libby:1978qf}.   Adding up all
cut diagrams is the same as summing diagrams of $A$ and then taking $|A|^2$.
If we want to identify the sources of IR enhancements,
we only need to find the set of possible classical processes that a diagram like this can represent.
\begin{figure}
\centerline{ \figscale{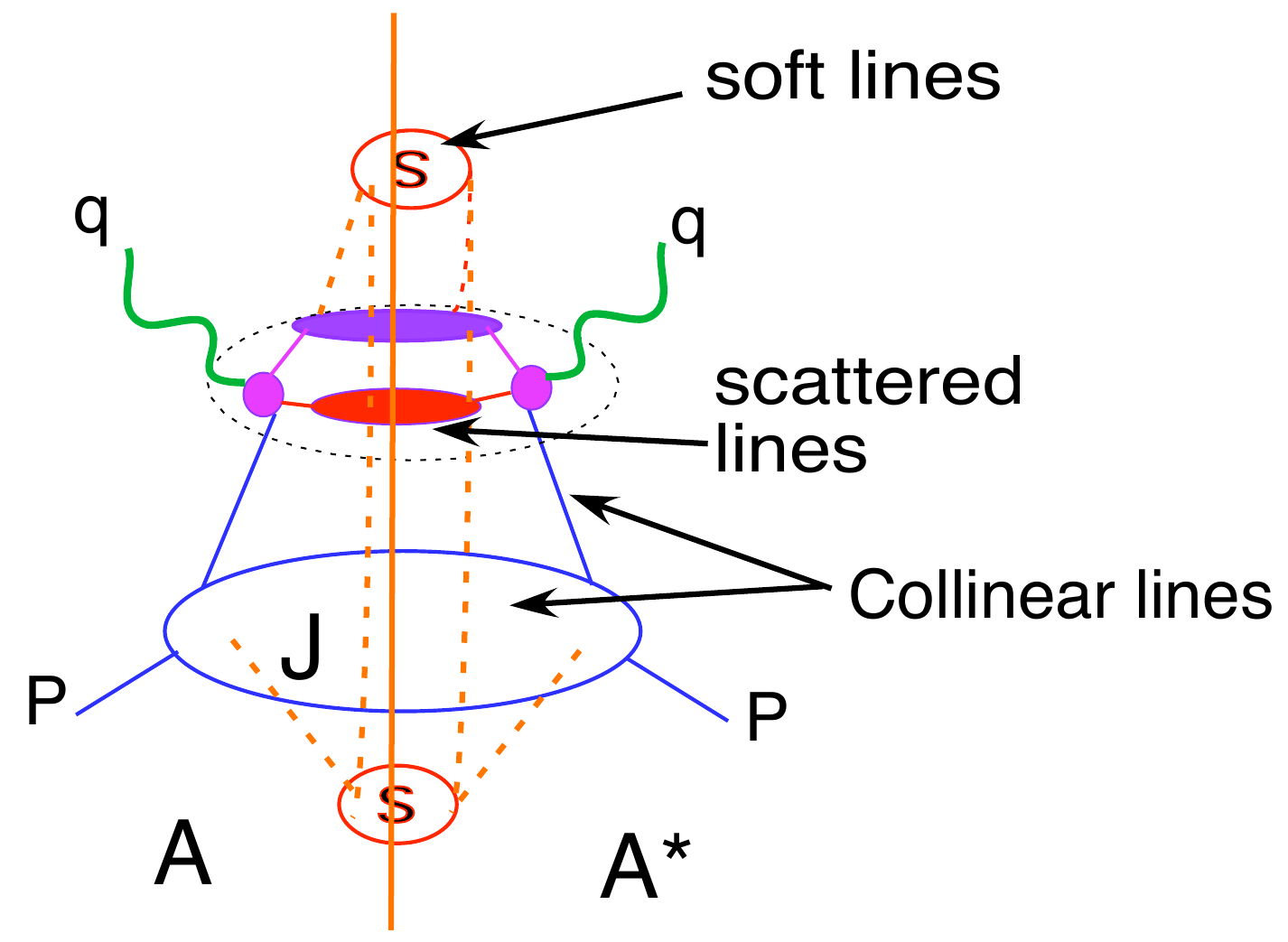}{6cm} \quad \quad \figscale{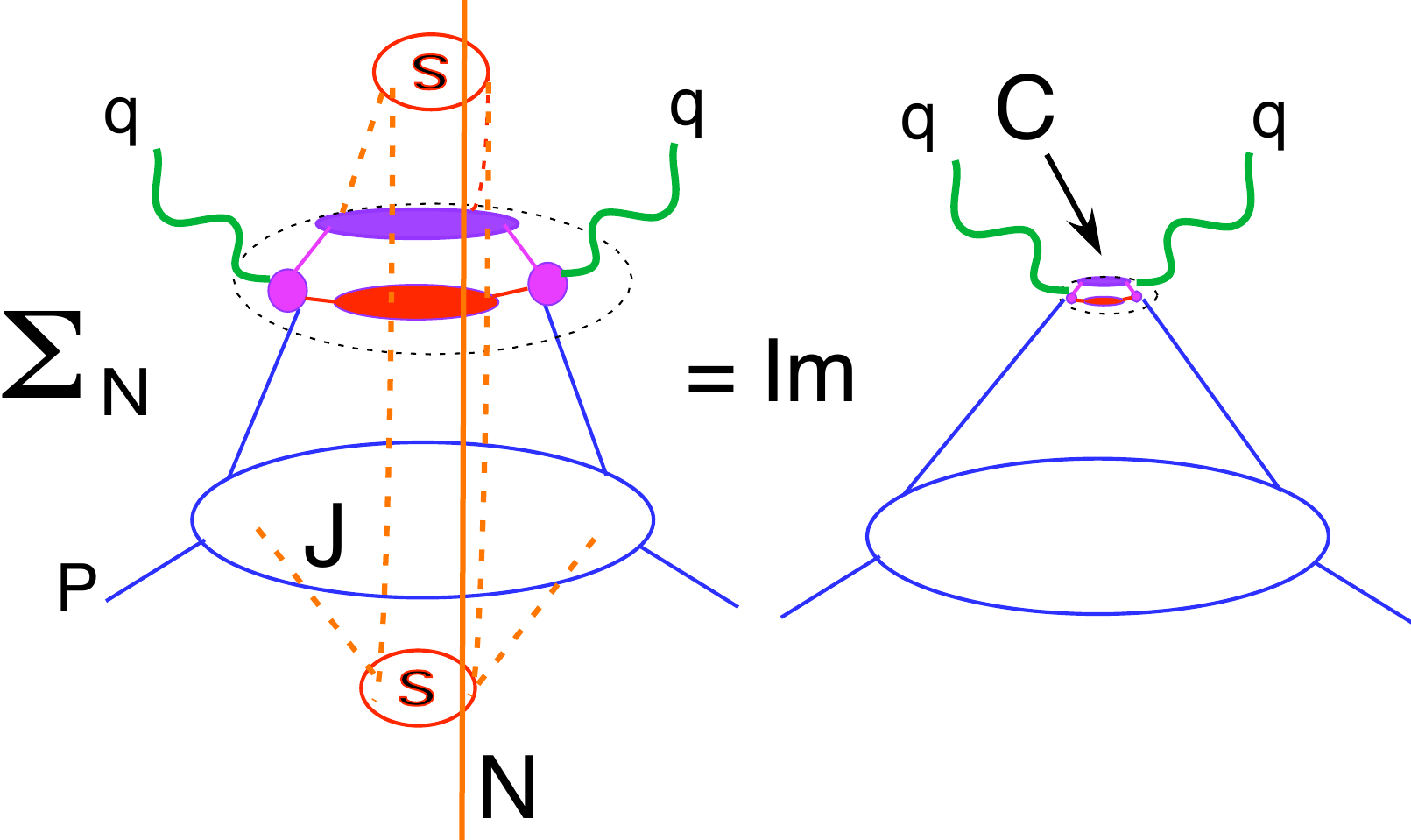}{6cm} } \vspace{4mm}
\centerline{(a) \hspace{8cm} (b) }
\caption{Cartoon of the proof of factorization in DIS. \label{fig:dis-fact}}
\end{figure}

In fact, there is really only one type of classical ``story" to be found:  the incoming particle (for us, quark or gluon)
of momentum $p$ splits into sets of collinear partons (we refer to these as the ``incoming jet"). Then {\it one} of these particles,
carrying momentum $xp$, with $0<x<1$,
 absorbs the virtual photon. Involving another parton of the incoming jet would cost additional powers of $Q$.\footnote{Strictly speaking, 
we must generalize these considerations for gauge theories in covariant gauges.  In a gauge theory like QCD we 
group with a physical parton those unphysical, parallel-moving gluons whose polarizations are proportional to the incoming jet
momentum.   After summing over  gauge invariant sets of diagrams, these gluons dress the parton  with a factorized Wilson line, as discussed in
connection with the operator definition of a parton distribution, Eq.\ (\ref{eq:phi-q-msbar}).}    
 The absorption of the photon results in the production of one or more sets of collinear partons,
 of total momentum $xp+q$, which recede from the incoming  jet and from each other  
at the speed of light, connected only by ``infinite wavelength soft" quanta.

Once this set of jets (and zero momentum
particles) crosses the final state from the amplitude, all lines begin to propagate backwards in time in the complex conjugate, with the
wide-angle jets reforming a single parton of the incoming jet, with the same momentum $xp$.  The system eventually combines itself back into the
initial-state parton of momentum $p$.   

The central observation is that jets of particles moving classically in different directions and
originating at a single point, can never meet again in the amplitude, only after 
crossing to the complex conjugate, where their time direction is reversed.   These consideration apply to all lines of
finite momentum in the cut diagram.   We can still ``dress" such lines with zero-momentum lines, which
we can think of as having infinite wave-length in the space-time picture of scattering.   For massless partons,
and at leading power in $Q$, this is the {\it only} kind of classical story DIS has to tell.

It is natural here to mention a systematic approach to these momentum configurations \cite{Becher:2014oda,Stewart:2009yx}
Soft collinear effective theory (SCET) builds this structure into calculations by
isolating the parts of the full QCD Lagrangian that correspond to different jets and
soft particles.    SCET organizes calculations that are equivalent to full QCD when factorization applies.

The picture in Fig.\ \ref{fig:dis-fact}a is suggestive of factorization, but it still has infrared sensitivity associated
with its outgoing jets, all emerging from a single point in the amplitude, and then rejoining at a comparable point in the
complex conjugate.   

At this stage, we are ready to use the ``inclusive" part of DIS.  Because we sum over all final states with the
same external particles (parton and virtual photon), we can use the optical theorem, which states that
the total cross section for a given initial state equals the imaginary part of that state's forward scattering amplitude, as shown
in Fig. \ref{fig:dis-fact}b.  In 
the forward amplitude, no classical processes involving jets in any direction other than the incoming direction
are possible.   Scattered partons would have to rescatter, but without the cut, which reverses their directions in time,
they can't rescatter, because
they would be moving away from each other at the speed of light.
All interactions after the hard scattering collapse
to a ``short-distance" function, labelled $C$ on the right of Fig.\ \ref{fig:dis-fact}b.   This
``coefficient function" then depends only on $xp$ and $q$.
A systematic analysis \cite{Collins:2011zzd,Sterman:1995fz} shows, in
particular, that long wavelength soft gluons in subdiagram  $S$ of Fig.\ \ref{fig:dis-fact}a can't resolve the ``tiny" remnant of the final states
that survives after the use of the optical theorem.  The  partons on each side of the short distance function $C(xp,q)$ now must have the same 
flavor, color and momentum fraction.   Most significantly, the general diagram on the lower part of the right-hand side of 
Fig.\ \ref{fig:dis-fact}b is equivalent order-by-order and point-by-point in momentum space to the diagrammatic
expansion of the quark distribution given in Eq.\ (\ref{eq:phi-q-msbar}), or its gluonic equivalent. 
These arguments can be applied to spin-dependent structure functions and the
corresponding polarized parton distributions \cite{Collins:1992xw,deFlorian:2009vb,Accardi:2012qut,AbdulKhalek:2021gbh}

\subsection{Evolution and the power of universality}

Equation (\ref{eq:F2-fact}), which expresses factorization, includes a dependence on the
value of the factorization scale, $\mu_F$.   But that scale is our choice, while the structure functions,
which are physical and can be measured, cannot depend on $\mu_F$.  As a result, we can write, for any
of the structure functions for the exchange of any electroweak vector $V=W^+,W^-,Z$ to hadron $h$,
\bea
\mu_F\frac{d}{d\mu_F} F^{Vh}_i(x,Q^2)  = \mu_F\frac{d}{d\mu_F} C^{Va}_i(x/z,Q/\mu_F,\as(\mu_F)) \otimes \phi_{a/h}(z,\mu_F,\as(\mu_F)
 = 0\, ,
\eea
where we have chosen our renormalization scale as $\mu_F$.  
This enables us to determine $\mu_F$ dependence by separation of variables.   That is, the $\mu_F$ dependence of
the coefficient function must cancel that of the parton distributions, and their variations with $\mu_F$ can depend only
on the variables held in common.   In covolution notation, we write
\bea
\mu_F\frac{d}{d\mu_F} C^{Va}_i  &=& -\, C^{Vc}_i\otimes P_{ca}\, ,
\nn\\
\mu_F\frac{d}{d\mu_F} \phi_{a/h} &=&  P_{ab} \otimes  \phi_{b/h}\, .
\label{eq:dC-dphi}
\eea
From these relations, we learn that we can compute the separation functions $P_{ab}$ from the infrared safe 
coefficient functions, so that the separation functions, the splitting functions or evolution kernels,
are themselves infrared safe.   This generalizes the calculation above leading to the quark-to-quark
evolution kernel, $P_{qq}$, Eq.\ (\ref{eq:Pqq}).   At arbitrary orders, the full set of parton distributions
mix, \cite{Altarelli:1977zs,Gribov:1972ri,Dokshitzer:1977sg}
\bea
\mu{d\over d\mu} \,\phi_{a/h}(\xi,\mu^2) = 
\sum_{b=q,\bar q,G}\int_{\xi}^1 {d\xi' \over \xi'}\
P_{ab}(\xi/\xi',\alpha_s(\mu))\,  \phi_{b/h}(\xi',\mu^2)\, .
\label{eq:evolution}
\eea
Splitting functions are known to three loops \cite{Vogt:2004mw,Moch:2004pa}, with four loops
on the horizon \cite{Moch:2023tdj}.

In summary, in contrast to the parton model distributions $\phi_{a/h}(x)$, which are
assumed to be independent of the scale at which the hadron, $h$ is probed,
scale dependence in QCD measures the change in parton distributions
as the factorization scale changes.  Of course, the cross sections we compute still depend on our 
choice of the factorization scale
through unknown higher orders in the coefficient functions.
  Nevertheless, to this approximation we can use distributions $\phi_{a/h}(x,Q_0^2)$ at scale $Q_0$ to determine
$\phi_{a/h}(x,Q^2)$ and hence $F^{Vh}_{1,2,3}(x,Q^2)$ for any $Q$.    
And indeed, evolution allows for any observable that can be factorized in this
manner to be computed.   This is how we found out that QCD is ``right".
Factorization applies to polarized as well as unpolarized cross sections,
and to the production of leptons and electroweak bosons, including
the influence of spin \cite{Collins:1992xw,STAR:2010xwx,PHENIX:2015ade}.
It is also how perturbative QCD can be used to make predictions based on new physics, so
long as the new processes are at short distances.   It therefore forms a foundation
for the application of QCD to the electroweak sectors of the Standard Model.

\section{Crossed EW-strong processes}
\label{sec:crossed}

Among the characteristics of relativistic quantum field theories are sets of relationships
between processes found by exchanging incoming particles with outgoing antiparticles.
We say that $a+b \to c+d$ and $a+\bar d \to c+\bar b$ are related by ``crossing".
We now turn to the  extension of the methods
for lepton-quark scattering to its crossed electroweak interactions,
leptonic annihilation to quark pairs and quark annihilation to lepton pairs..
These have been sources of discoveries within the Standard Model, and
are among the primary tools in searches for new physics beyond it.  

\subsection{Leptonic annihilation}

Applied to lepton-quark scattering, crossing leads to leptonic annihilation,
relating, for example, deep-inelastic processes like $p+e \to X + e$, and
single-particle inclusive cross sections like  $\bar e +  e \to   X + \bar p$.
The relationship is illustrated schematically in Fig.\ \ref{fig:1pi}, and it
is natural to apply parton model intuition to the crossed process.
\begin{figure}
\centerline{ \figscale{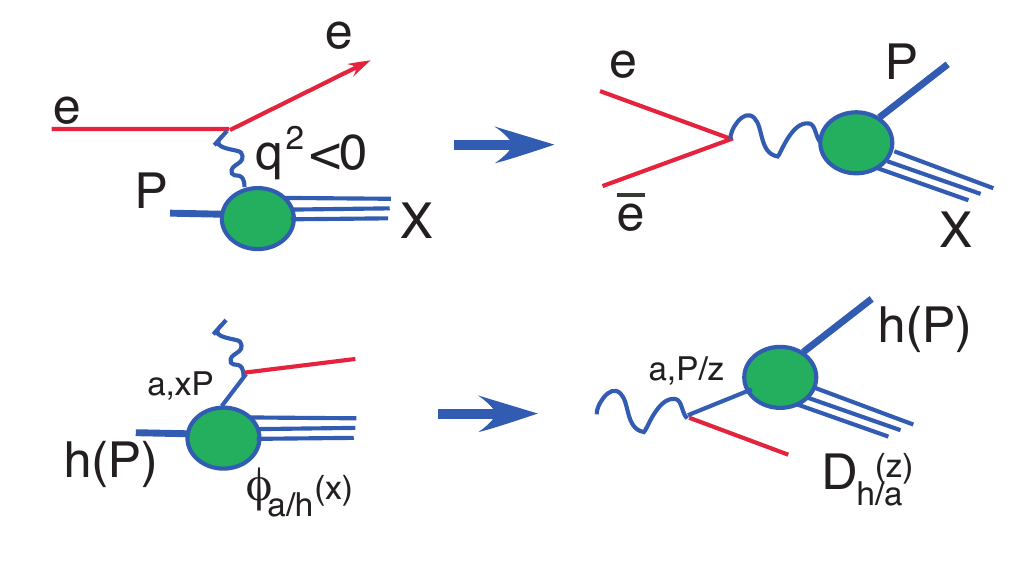}{8cm}}
\caption{Crossing from DIS and parton distributions to 1PI in leptonic
annihilation and fragmentation functions. 
The upper line relates the processes, the lower shows their parton model pictures.
\label{fig:1pi}}
\end{figure}

For the crossed processes, the underlying lowest-order electroweak reaction
is $e + \bar e \to q + \bar q$, with $q$ any of the quarks that can be produced
at the energy of the annihilating pair.   Then, taking the place of parton distributions,
we introduce fragmentation functions, which are
thought of as the probability distributions of ``hadrons in a parton". 
The parton model  relation for the single particle inclusive (1PI) cross section to produce a hadron of momentum $p_h$ 
in this process is given by analogy to the DIS expression, Eq.\ (\ref{eq:parton-sigma}),
\bea
&& \hspace{-30mm}
\frac{d\sigma^{\rm incl}_{{\rm e^+e^-}\to h+ X}(p_h,q)}{d^3p_h}  = \sum_a
 \int_0^1 dz\,  z^3\,\frac{d\sigma^{\rm el}_{{\rm e^+e^-}\to a+\bar a}(p_h/z,q)}{d^3p_h}\; D_{h/a}(z)\ ,
\label{eq:1pi-pm}
\eea
with the inclusive hadronic cross section on the left and the elastic parton-pair cross section on the right, where
$D_{h/a}(z)$ is the probability distribution of hadron $h$ with momentum $p_h$ resulting from the production of parton $a$ with momentum $p_h/z$.  Incorporating QCD, we can
formulate a factorized version of Eq.\ (\ref{eq:1pi-pm}), with corresponding evolution equations \cite{Mueller:1978xu}.
The ``fragmentation function", $D_{h/a}(z)$ is universal, and can also describe single-particle cross sections in DIS or hadron-hadron scattering \cite{Nayak:2005rt}.

We find a heuristic justification for the parton picture of Eq.\ (\ref{eq:1pi-pm}) from time dilation: the formation of hadron $h(p)$ from parton $a(p/z)$
takes a fixed time $\tau_0$ in the rest frame of $a$, but much longer
in the CM frame.  This fragmentation thus decouples 
from $d\sigma^{\rm el}_{{\rm e^+e^-}\to a+\bar a}/d^3p$, and is independent of $q$.   The sole dependence on $Q^2$ is in the 
elastic process.   We can introduce a structure function description of single-particle inclusive cross sections
in much the same way as for DIS, and derive a corresponding scaling property.  

A very suggestive feature of the parton model picture for 1PI cross sections is that the direction of the hadron follows the direction 
of the parton,\footnote{Suggested as a possibility in Ref.\ \cite{Bjorken:1969wi}}  just as the parton's momentum is collinear to the incoming hadron's momentum in DIS.   Implicitly then, in the parton
picture, hadrons emerge predominantly in the directions of the original pair $q,\bar q$.   In the center-of-mass frame
they are back to back.  Since the angular distribution of the lepton to parton annihilation process depends on the spin of the 
partons, the distributions of the hadrons should directly reflect this spin.   For spin-1/2 partons, the quarks of our world, the 
cross section is
\bea
\frac{d\sigma^{\rm el}_{ e\bar{e} \ra f \bar{f}}(k_1,k_2) }{d\Omega_{\rm cn}} 
&=&
{3Q_f^2 \alpha^2 \over 4Q^2}\, \left (1 + \cos^2\theta \right)\, ,
\label{eq:epem-LO}
\eea
with $Q^2=(k_1+k_2)^2$, and $\theta$ the angle between the electron and the quark (or antiquark) in the center of mass.  
Here $Q_f=e_f/e$ is the electric charge of quark $f$ in units of the positron charge (2/3 or -1/3) and the factor of $3=N_{\rm color}$ reflects the equal electroweak couplings of the three colors in the fundamental representation of QCD.   This angular distribution
is the leptonic annihilation analog of the Callan-Gross relation Eq.\ (\ref{eq:C-G-relation}) in DIS.

The fragmentation picture suggests that almost all hadrons are aligned along
parton directions, implying that most hadrons come out together as ``jets", 
following the $1+\cos^2\theta$ distribution relative to the incoming electron.  
And this is what happens \cite{Hanson:1975fe}, and not only in leptonic annihilation.
The key to quantifying this expectation is to follow the flow of energy carried by hadrons.

In fact, we can formulate jet cross sections that are calculable in perturbation theory.   All we have to
do is identify observables that are unchanged when a parton of zero energy is emitted and when 
a massless particle decays into two massless particles moving in the same direction \cite{Sterman:1979uw}.   Such observables
can be built from energy flow operators, ${\cal E}(\Omega)$, with $\Omega$ a direction on the sphere, defined by their actions on states,
\cite{Sterman:1975xv,Sterman:1977wj,Basham:1978bw,Sveshnikov:1995vi,Korchemsky:1997sy,Larkoski:2013eya,Lee:2022uwt}
\bea
{\cal E}(\Omega)\, | \{ k_i\}\rangle\ =\ \sum_i k_i^0\, \delta^2(\Omega-\Omega_i) | \{ k_i\}\rangle\, .
\eea
Cross sections constructed from such operators, which respect the flow of energy, provide a large set of infrared safe cross sections
in leptonic annihilation processes, and in the decay of electroweak bosons \cite{Butterworth:2008iy}.  Among these are various cross sections for hadronic jets and event shapes \cite{Larkoski:2017jix}.

\subsection{Drell-Yan and allied gateways}

The application of factorization that is most relevant to discovery and now precision studies of the electroweak sector is the Drell-Yan process.  Historically, it was an extension of parton model ideas to hadron-hadron scattering.   Just as a lepton pair can annihilate to a pair of quarks through electroweak interactions, so can a quark and antiquark, one from each of a colliding pair of protons, say, annihilate to produce a pair of leptons at high invariant mass, whether as part of a continuum or as a enhancement \cite{Smith:1983aa} at the mass
 of an electroweak boson \cite{Drell:1970wh,CDF:2022hxs,CMS:2024lrd}.   This of course was how the W${}^\pm$ and Z were first discovered \cite{UA1:1983crd,UA1:1983mne}, and extending the concept to gluon pair annihilation, the Higgs boson as well \cite{CMS:2012qbp,ATLAS:2012yve}.  To the extent the production cross sections of electroweak states are understood, their properties can be studied, including their masses, decays and couplings.   Here, we shall simply review the main results for cross sections, and sketch some of their justification.

We consider  the production
of a real or virtual electroweak boson $B$, of momentum $q$ with invariant mass $Q=\sqrt{q^2}$
and rapidity $\eta=(1/2)\ln (q\cdot p_1/q\cdot p_2)$
 in the collision of hadrons $H_1$ and $H_2$ , which we take in the center of mass plus and minus directions, respectively.
 The boson is taken to have momentum components 
 \bea
 q^+ &=& \frac{q\cdot p_2}{\sqrt{s/2}} \equiv x_a\sqrt{s/2}\, ,
  \nn\\
  q^- &=& \frac{q\cdot p_1}{\sqrt{s/2}} \equiv x_b\sqrt{s/2} \, .
  \label{eq:qpm-def}
  \eea
 In these terms, the cross section be written in factorized form as \cite{Collins:1989gx,Collins:2011zzd}
\bea
\frac{d\sigma_{\rm H_1H_2 \to B+X} }{dQ^2 d\eta}&=& 
\sum_{a,b} \int_0^1 d\xi_a\, d\xi_b\,
H_{ab\to B+X}\left(\frac{x_a}{\xi_a},\frac{x_b}{\xi_b}, Q,\mu,\alpha_s(\mu)\right)\,
 \phi_{a/H_1}(\xi_a,\mu)\,\phi_{b/H_2}(\xi_b,\mu) \, ,
 \nn\\
 \label{eq:hh-fact}
\eea
where we sum (implicitly) over all {\it hadronic} final states $X$ that result from the annihilation of quarks. 
We have taken the factorization and renormalization scales equal for this discussion.  As shown,
this is the cross section at measured mass ($Q$) and rapidity ($\eta$) of the boson that is produced
by the quark pair's annihilation.  In this form, we sum as well over the decays of boson $B$; these may be
either leptonic or hadronic.  Measurements of these features of the final state do not affect the basic form of the factorization
for unpolarized initial states.

In Eq.\ (\ref{eq:hh-fact}), $H_{ab\to B+X}$ is an infrared safe ``hard" function, with an expansion
in the strong coupling that begins with the elastic (Born) cross section for 
quark-antiquark annihilation (here through a photon) to the boson in question,
 \bea
 H^{(0)}_{a\bar b\to B+X} = \delta_{b \bar a} e_a^2 \frac{4\pi \alpha^2}{9Q^4}\, \delta\left( 1 - \frac{x_a}{\xi_a}\right)\, \delta\left( 1 - \frac{x_b}{\xi_b}\right)\, ,
\label{eq:dy-born}
\eea
with $e_a$ the charge of quark $a$ in units of positron charge.   Most importantly, in Eq.\ (\ref{eq:hh-fact}) 
the functions $\phi_{a/H}(\xi,\mu_F)$ are the same parton distributions as in deep inelastic scattering,
with the same dependence on their factorization scale.  

 At the time of the discovery of the $W$ and $Z$ bosons,
the hard scattering function for the inclusive cross secction was known to one loop \cite{Altarelli:1979ub,Altarelli:1984pt}. 
  In the intervening time, vector boson and 
Higgs QCD inclusive and some differential cross sections have been calculated at two and even three loops \cite{Dawson:1990zj,Spira:1995rr,Campbell:1999ah,Ravindran:2002dc,Harlander:2002wh,Ravindran:2003um,Anastasiou:2003ds,Catani:2009sm,Anastasiou:2015vya,Mistlberger:2018etf}.

The proof of collinear factorization with universal parton distributions relies on techniques and observations similar to those in
arguments for factorization in DIS, with additional considerations since there are two hadrons in the
initial state.   As in DIS, all long-distance dependence is associated with free, classical propagation of
on-shell particles.  Now there are two ``jets" of incoming particles, sequences of virtual states emerging from the two
particles of the initial state \cite{Collins:1983ju,Bodwin:1984hc,Collins:1985ue,Collins:1988ig,Collins:2011zzd}.   When a quark from one incoming jet annihilates with an antiquark
from the other to form boson $B$, other high energy particles may be produced, against which
$B$ recoils.  Because we are inclusive in hadronic final states, however, the unitarity of QCD
ensures that all long-distance interactions of these particles cancel, and they appear 
as corrections to the hard function $H_{ab\to B+X}$.    This ensures the cancellation
of effects from the interactions of partons after the annihilation of the quark and antiquark. 
These considerations are essentially
equivalent to those in DIS, as is the treatment of unphysical collinear gluons in the directions
of the incoming particles.  
Compared to DIS, however, the treatment of soft, long wavelength, lines
is more complex, because such lines can connect the two incoming jets, regardless of the 
short-distance process.

For the soft lines, arguments for factorization rely on causality as well as unitarity, and
hold with universal parton distributions only when the initial state hadrons approach each other
at nearly the speed of light.   In this limit, partons in one hadron cannot communicate with each
other before the scattering actually occurs.  Very much as for the electron in DIS, each hadron
``sees" the other as frozen in one of its virtual states.  At lower velocities, the quarks in one hadron
might, for instance, have time to attract the antiquarks in the other before the collision.   Such an effect
would obviously conflict with factorization with universal parton distributions.   

In the collinear-factorized cross section of Eq.\ (\ref{eq:hh-fact}) we do not fix the transverse momentum $Q_T$ of boson $B$.
The inclusive cross section is dominated by the region $Q_T\ll Q$.
In fact, if $Q_T$ is ``large",  of order $Q$, the same factorized form holds, with
explicit $Q_T$ dependence in the hard function, $H$.   In that case, we can
calculate this cross section with boson $B$ recoiling against one or more quark and/or gluon jets.
When $Q_T$ becomes small compared to the boson mass $Q$, however, we encounter logarithms
of the ratio $Q/Q_T$, up to the power $\as^n \ln^{2n-1}(Q/Q_T)$, with an overall power of $1/Q^2_T$.   
This phenomenon is similar to the behavior of the DIS coefficient function,
where we encountered the generalized function $[\ln(1-x)/1-x]_+$ 
in the order $\as$ coefficient function, Eq.\ (\ref{eq:C2-1}).  
Just as the DIS coefficient is finite through cancellations of real and virtual radiation,
so the inclusive Drell-Yan cross section is calculable through analogous cancellations. 
Strikingly, for electroweak boson 
production, we are also able to control, ``resum", the leading behavior at {\it measured} $Q_T$,
to all orders in perturbation theory.
To do so, we must generalize collinear factorization itself.

\subsection{$Q_T$ factorization and resummation}

The 
Drell-Yan process enjoys a further factorization, which makes possible 
the calculation of its transverse momentum dependence at high energies even down to vanishing $Q_T$.   This is
perhaps the best example of how factorization properties can lead to
the resummation of large corrections at all orders in perturbation theory.\footnote{A very
similar procedure of ``threshold reummation" applies to the $x\to 1$ enhancements in DIS and inclusive
electroweak boson production \cite{Sterman:1986aj,Catani:1989ne,Kulesza:2002rh}.
The beautry of $Q_T$ resummation is that it applies to a direct observable.}
We close our discussion with a slightly simplified treatment, which 
will illustrate the main points, including the connection between factorization
and evolution in a more general context \cite{Contopanagos:1996nh}.

The so-called $Q_T$-factorization, which generalizes collinear factorization in Eq.\ (\ref{eq:hh-fact}),
applies to the part of the cross section that is singular as $1/Q_T^2$ 
as $Q_T\to 0$, in terms of a new kind of parton distribution \cite{Parisi:1979se,Curci:1979sk,Collins:1981uk,Collins:1984kg},
\bea
  {d\sigma_{H_1,H_2\rightarrow B+X} \over dQ^2\, d\eta\, d^2Q_T}
&=& \ \sum_a
H(x_ap_1,x_bp_2,Q, n)_{a\bar{a}\rightarrow B } \
 \nn\\
&\ & \times \int d^2{\bf k}_{1T}\, d^2{\bf k}_{2T}\, d^2{\bf k}_{sT}\, \delta^2\left ( {\bf Q}_T-{\bf k}_{1T}-k_{2T}-{\bf k}_{sT}\right)
\nn\\
&\ & \quad  \times\ {\cal P}_{a/H_1}(x_a, p_1\cdot n,{\bf k}_{1T})\, 
 {\cal P}_{\bar{a}/H_2} (x_b,p_2\cdot n,{\bf k}_{2T} )\  
 U_{a \bar{a}}({\bf k}_{sT},n)\, .
\label{eq:qt-fact}
\eea
Here, the ${\cal P}_{i/H_i}(x_i,p_i\cdot n,{\bf k}_{iT})$ are transverse momentum-dependent distributions (TMDs), 
which describe the transverse momentum ${\bf k}_{iT}$ given up by the parton at a given parton fractional momentum $x_i$ \cite{Collins:2011zzd,Boussarie:2023izj}. 
We also  introduce $U$, a ``soft function" for wide-angle radiation, not associated with the directions of the incoming partons.
The transverse momentum of the boson is the negative of the sum of these gluon emissions.   
Proofs of this extended factorization have been given in Refs.\ \cite{Collins:2011zzd} and \cite{Laenen:2000ij}.
Its validity depends strongly on the nature of the cross section, involving the observation of 
only leptons or electroweak bosons in the final state \cite{Collins:2007nk,Collins:2007jp}.   Measurements of hadronic pairs become sensitive
to soft interactions that take place after the pair annihilation event, and in general spoil the form
given in Eq.\ (\ref{eq:qt-fact}).

We will not give detailed, matrix element definitions of the functions in Eq.\ (\ref{eq:qt-fact}),
analogous to Eq.\ (\ref{eq:phi-q-msbar}) for
collinear distributions.
These may be found in \cite{Collins:2011zzd,Boussarie:2023izj}.  We note, however, their dependence
on the additional vector $n^\mu$.   Very roughly, $n^\mu$ apportions particles in the final state in terms of their momenta, $k$.
For example, if the inner product $k\cdot p_a$ is smaller than $k\cdot p_b$ and $|k\cdot n|$, we 
associate particle $k$ with ${\cal P}_{a/N}$,  and similarly for the other two functions.
As we shall see, the choice of the precise direction for $n^\mu$ plays a role similar to the factorization scale in
collinear evolution.
 
Starting from the factorized expression (\ref{eq:qt-fact}), which is a convolution
in transverse momenta, it is natural to 
examine this expression as the Fourier transform of the simple product of the TMDs and soft radiation function in 
the conjugate ``impact parameter" space, labelled by a two-dimensional vector, ${\bf b}$,
\bea
\int d^2 Q_T\, e^{-i{\bf b}\cdot {\bf Q} }\; {d\sigma_{H_1H_2\rightarrow B+X} \over dQ^2\, d\eta\, d^2Q_T }
&=& \sum_a  H(\xi_1p_1,\xi_2p_2,Q,n)_{a\bar{a}\rightarrow B } 
   \nn \\
 &\ & \hspace{-45mm}
 \times 
 \tilde {\cal P}_{a/H_1}(x_a,p_1\cdot n/\mu_R,b\mu_R)\, 
 \tilde{\cal P}_{\bar{a}/H_2}(x_b,p_2\cdot n/\mu_R,b\mu_R )\  \tilde U_{a \bar{a}}(b\mu_R,n)\, .
\label{eq:b-space}
\eea
Now we can resum by separating variables, in the manner we used to derive 
 evolution above in DIS.
The cross section, of course, is independent of both the renormalization scale $\mu_R$ and of $n^\mu$, giving two equations,
\bea
 \mu {d\sigma \over  d\mu} = 0\, , \hspace{10mm} n^\alpha{d\sigma \over dn^\alpha}=0\, ,
\eea
where here and below we drop the subscript $R$ on the renormalziation scale.
In the following, we exploit these conditions using a method due to Collins and Soper, \cite{Collins:1981uk} and Sen \cite{Sen:1981sd}.

The dependence on $n^\mu$ in the TMDs must cancel dependence in (UV) function $H$ and (IR) function $\tilde U$, which depend on different variables,
\bea
p_i\cdot n\, \frac{\partial}{\partial p\cdot n}\, \ln\, \tilde{\cal P}_{c/H_i}\left (x,{p_i\cdot n \over \mu}, b\mu,\as(\mu) \right ) 
= \frac{1}{2}\, G_c\left ({p_i\cdot n \over \mu},\as(\mu) \right) + \frac{1}{2}\,  K_c(b\mu,\as(\mu))\,)\, .
\label{eq:P-derivative}
\eea
Here,
the function $G_c$ matches changes in  the hard scattering function $H$, and $K_c$ matches soft function $\tilde U$.  
Next, we use that the functions $\tilde{\cal P}_{c/H}$ are rendered ultraviolet finite by
the standard multiplicative renormalization (the transverse
integrals that define the TMDs are constrained by the toal $Q_T$). 
The right-hand side of (\ref{eq:P-derivative}) is therefore renormalization-scale independent \cite{Collins:1981uk,Sen:1981sd},
\bea
\mu\, \frac{\partial}{\partial \mu} \left[\, G_c \left ({p_i\cdot n \over \mu},\as(\mu) \right) + K_c(b\mu,\as(\mu))\, \right] = 0\, .
\label{eq:K+G}
\eea
Once again, separating variables, we find a new function for parton $c$, which depends on the coupling only,
\bea
\mu\, \frac{\partial}{\partial \mu} \, G_c \left ({p_i\cdot n \over \mu},\as(\mu) \right)\
=\ \gamma_{K,c}(\alpha_s(\mu))\ =\ -\
\mu\, \frac{\partial}{\partial \mu}\, 
K_c(b\mu,\as(\mu))\, .
\label{eq:A-def}
\eea
Using this key relation to shift the renormalization scale in function $K_c$ from $\mu$ to $1/b$ 
and $G_c$ from $\mu$ to $p\cdot n$ then gives for the variation of the TMD, Eq.\ (\ref{eq:P-derivative}),
\bea
p\cdot n\, \frac{\partial}{\partial p\cdot n}\, \ln\, \tilde{\cal P}_{c/H_i} \left (x,{p_i\cdot n \over \mu}, b\mu,\as(\mu) \right ) 
&=& \frac{1}{2}\,  G_c \left ({1},\as(p_i\cdot n) \right) + \frac{1}{2}\,  K_c(1,\as(1/b))
\nn\\
&\ & \quad
- \frac{1}{2}\,  \int_{1/b}^{p_i\cdot n} \frac{d\mu'}{\mu'}\, \gamma_{K,c}(\alpha_s(\mu'))\, .
\label{eq:new-eqn}
\eea
On the right, the integral produces logarithms of $b \times p_i\cdot n $. 
The factor $1/2$ is included to match the notation of \cite{Collins:1984kg}.
 We next integrate $p\cdot n$ down to $1/b$,
and finally set $\mu=1/b$ in all functions.
The result generates double logarithms in $b$ at each order, a sum in $\as^n\ln^{2n}bQ$.
Transformed back to $Q_T$, and setting $p_1\cdot n=p_2\cdot n=Q$, we get all terms of the form $\ln^pQ^2/Q_T^2$, with $p=1 \dots 2n-1$, from the expression
\cite{Parisi:1979se,Curci:1979sk,Collins:1981uk,Collins:1984kg},
\bea
      {d\sigma_{H_1H_2\rightarrow QX}(Q,b) \over dQ^2\, d\eta\, d^2Q_T }
       &=& 
\sum_a\, H_{a\bar{a}}(\as(Q^2)) \,
    \int \frac{d^2b}{(2\pi )^2} \, e^{ i {\bf Q}_T\cdot {\bf b}}\, 
 \exp \left[ E_{a\bar{a}}^{\rm PT} (b,Q,\mu) \right ]
\nonumber\\
&\ & \hspace{-10mm}  \times\ 
 {\cal P}_{a/H_1}(x_a,1,1,\as(1/b))\, 
 {\cal P}_{\bar{a}/H_2}(x_b,1,1,\as(1/b) )\  U_{a \bar{a}}(1,n,\as(1/b))\, ,
\label{eq:resum-qt}
\eea
where we have exhibited dependence on the strong coupling, now at scale $1/b$.
The ``double-logarithmic" exponential  links large and low virtuality, \cite{Collins:1984kg}
\bea
E_{a\bar{a}}^{\rm PT} = - 
\int_{1/b^2}^{Q^2} \, {d \mu^2 \over \mu^2}\;
\left[ A_a(\as(\mu))\,
\ln\left( {Q^2 \over \mu^2} \right)\, + B_a(\as(\mu))\right]  \, ,
\label{eq:exponent}
\eea
where leading logarithms are generated from the term with $A_q=(1/2)[\gamma_K+\beta(g) \partial K/ \partial g]=C_F+{\cal O}(\as^2)$.
The precise value of  $B_q= - [K(1,\as(Q))+G(1,\as(Q))]$ depends on choices for the organization of non-leading logarithms.
This resummed expression shows an exponential suppression as  soon as the
impact parameter $b$ becomes much larger than $1/Q$, which asymptotically forces the integral into the 
region of $b$ small enough that $\as(1/b)$ remains perturbative.   The physical basis of this suppression is that in
a hard annihilation process, the termination of the two color charges always results in gluon emission,
and to a set of final states in which the transverse momentum carried by those gluons is highly constrained.

Nonperturbative dependence
in the TMDs $\tilde {\cal P}_{a/H}$  for larger $b\sim  \Lambda_{\rm QCD}$ does remain of interest even at collider energies.
With modest, and factorized, nonperturbative input, however,
this result can fit the low-$Q_T$ data for electroweak vector and Higgs production with real precision \cite{Bozzi:2005wk,Isaacson:2023iui,ATLAS:2014alx,CDF:2022hxs}.
This formalism provides us with a window into  the virtual states
that are at the heart of evolution.

\section{Conclusion}

We have discussed some of the basis for studies linking the strong and electroweak interactions of the Standard Model,
concentrating on the classic examples that grew out of the parton model.  The parton model served to lay much of the physical
groundwork for the discovery of QCD.  With asymptotic freedom, higher-order corrections to 
the basic processes between strongly interacting hadrons and electroweak bosons can be
understood systematically.   This review has illustrated the impact of these developments
in a few fundamental processes, but there are many more that reflect electroweak interactions,
including direct photon, boosted electroweak boson, and associated Higgs production.
The universality of collinear parton distributions and the technique of 
evolution make possible predictions at future colliders and for rare processes, including those 
due to new physics.   

\subsection{Acknowledgements}

The author thanks the editors of the review for their invitation to contribute.  This work was supported in part by the National Science Foundation, award PHY-2210533.

\end{document}